\documentstyle[sprocl]{article}

\input{psfig}

\newcommand{\lsim}{\raisebox{-0.6ex}{
                        $\stackrel{\textstyle <}{\textstyle\sim}$}}

\newcommand{\beq}{\begin{equation}}
\newcommand{\eeq}{\end{equation}}

\begin{document}

\title{POWER COUNTING FOR AND SYMMETRIES OF THE EFFECTIVE FIELD THEORY FOR $NN$
INTERACTIONS}

\author{MARK B. WISE}
\address{California Institute of Technology, Pasadena\\ CA  91125, USA \\
E-mail: wise@theory.caltech.edu}

\maketitle\abstracts{The appropriate power counting for the effective field
theory of
$NN$ interactions is reviewed.  It is more subtle than in most effective field
theories since in the limit that the $S$-wave $NN$ scattering lengths go to
infinity it is governed by a nontrivial fixed point.  The leading two body
terms in the effective field theory for nucleon self interactions are scale
invariant and invariant under Wigner SU(4) spin-isospin symmetry in this limit.
 Higher body terms with no derivatives (i.e., three and four body terms) are
automatically invariant under Wigner symmetry.}

\section{Basics}

Consider $N({\bf p}) N(- {\bf p}) \rightarrow N ({\bf p}') N (-{\bf p}')$
scattering in the $^1S_0$ channel.  Since the spins of the two nucleons are
combined anti-symmetrically Fermi statistics implies that this channel is $I =
1$ (similarly the $^3S_1$ and $^3D_1$ channels are $I = 0$).  The energy $E =
p^2/M = p^{\prime 2}/M$ where $p^{(\prime)} = |{\bf p}^{(\prime)}|$ and the
scattering matrix, $S$, is related to the scattering amplitude ${\cal A}$ by $S
= 1 + i Mp {\cal A} /2\pi$.  Since $S = e^{2i\delta}$, where $\delta$ is the
phase shift,
\begin{equation}\label{1}
{\cal A}^{(^1S_0)} = \frac{4\pi}{M} \frac{1}{p\cot \delta^{(^1S_0)} - ip},
\end{equation}
where $M$ is the nucleon mass.  For $p< m_\pi/2$ the quantity $p\cot\delta$ can
be expanded in a power series in $p^2$
\begin{equation}\label{2}
p\cot\delta^{(^1S_0)} = - \frac{1}{a^{(^1S_0)}} + \frac{1}{2} r_0^{(^1S_0)} p^2
+ \ldots ,
\end{equation}
when $a$ is called the scattering length and $r_0$ is called the effective
range.  The scattering length in the $^1S_0$ channel is very
large,~\cite{burcham}
$a^{(^1S_0)} \simeq - 23.7$ fm or $1/a^{(^1S_0)} \simeq - 8.3$ MeV.  On the
other hand the nuclear potential is characterized by a momentum scale $\Lambda
\sim 200$ MeV.  The smallness of $|1/a^{(^1S_0)}|$ compared with this scale is
the result of an accidental cancellation which causes a state in the spectrum
to be very near zero binding energy.  ($a \rightarrow - \infty$ as a scattering
state approaches zero energy and $a \rightarrow \infty$ as a bound
state approaches zero binding energy.)  Neglecting the small $^3S_1-
^3D_1$ mixing, formulas analogous to eqs.~(\ref{1}) and~(\ref{2}) hold in the
$^3S_1$ channel.  The scattering length is also large in that
case,~\cite{burcham} $a^{(^3S_1)}
\simeq 5.4$ fm or $1/a^{(^3S_1)} \simeq 36$ MeV.  The bound state in this
channel that is near zero binding energy is the deuteron.

\section{Expansions of ${\cal A}$}

The simplest expansion of ${\cal A}$ is a momentum expansion.  This is
analogous to what is done in standard applications of effective field theory,
e.g., chiral perturbation theory for $\pi\pi$ scattering.  For $NN$ scattering
in the $s =~ ^1\! S_0$ or $^3S_1$ channels,
\[
{\cal A}^{(s)} = \frac{4\pi}{M} \frac{1}{\left[- 1/a^{(s)} + \frac{1}{2}
r^{(s)}_0p^2 + \ldots - ip\right]}\]
\begin{equation}\label{3}
= - \frac{4\pi}{M} a^{(s)} \left\{1 - i a^{(s)} p + \left(\frac{a^{(s)}
r_0^{(s)}}{2} - a^{(s)^{2}}\right) p^2 + \ldots\right\}.
\end{equation}
If $a^{(s)}$ was its natural size (i.e., $a^{(s)} \sim 1/\Lambda$) this would
be the appropriate expansion to perform.  However, in nature the $S$-wave $NN$
scattering lengths are very large and the expansion above is only valid in the
small region of momentum $p\lsim |1/a^{(s)}| \ll 1/\Lambda.$  Since the
underlying physics is set by $m_\pi$ and $\Lambda_{QCD}$ there should be an
expansion in $p/\Lambda$ that is valid even when $p \gg |1/a^{(s)}|$.  It is
not difficult to deduce what this expansion is.  In Eq.~(\ref{3}) keep
$-1/a^{(s)} - ip$ in the denominator and expand in the remaining terms.
This yields
\begin{equation}\label{4}
{\cal A}^{(s)} = - \frac{4\pi}{M} \frac{1}{(1/a^{(s)} + ip)} \left[1 +
\frac{r_0^{(s)} p^2/2}{(1/a^{(s)} + ip)} +\ldots\right].
\end{equation}
Now ${\cal A}^{(s)} = \sum_{n = - 1}^{\infty} {\cal A}_n^{(s)}$, where
${\cal A}_n^{(s)} \sim {\cal O} (p^n)$.  This is the appropriate
expansion in the case where the scattering lengths are large.  It has the
unusual property that the leading term is order $p^{-1}$.

\section{Effective Field Theory Without Pions}

The effective field theory with the pions integrated out contains only nucleon
fields, $N = \left({p\atop n}\right)$, and we expect that the lowest dimension
operators
will be the most important ones.  The Lagrange density is written as,
${\cal L} = {\cal L}_1 + {\cal L}_2 + \ldots,$ where ${\cal L}_n$
contains $n$-body operators.  The one and two body terms are:
\begin{equation}\label{5}
{\cal L}_1 = N^\dagger \left[i \partial_t + \frac{\vec \nabla^2}{2M}\right]
N + \ldots,
\end{equation}
\begin{equation}\label{6}
{\cal L}_2 = - \sum_s C_0^{(s)} (N^T P_i^{(s)} N)^\dagger (N^T P_i^{(s)} N)
+ \ldots .
\end{equation}
Here $s = ~^1S_0$ or $^3S_1$, the ellipses denote higher dimension operators
and $P_i^{(s)}$ are the spin-isospin projectors
\begin{equation}\label{7}
P_i^{(^1S_0)} = \left(\frac{\sigma_2 \tau_2 \tau_i}{\sqrt{8}}\right), P_i^{(^3
S_1)} = \left(\frac{\sigma_2 \sigma_i \tau_2}{\sqrt{8}}\right),
\end{equation}
where the Pauli matrices $\sigma_i$ act in spin space and the Pauli matrices
$\tau_i$ act in isospin space.

\begin{figure}[t!]
\centerline{\psfig{figure=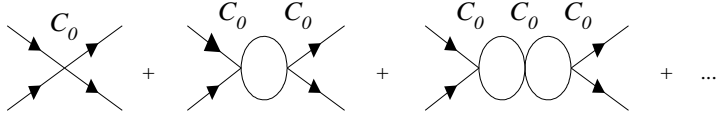,width=17.0cm}}
\caption[1]{The leading contribution to $NN$ scattering.}
\label{fig_C00}
\end{figure}
Neglecting higher dimension operators the scattering amplitudes in the $^1S_0$
and $^3 S_1$ channels come from the sum of bubble-type Feynman diagrams shown
in Fig. 1.  Each bubble is linearly divergent in the ultraviolet so the
coefficients $C_0^{(s)}$ depend on the regulator and subtraction scheme
adopted.  We use dimensional regularization and start with  minimal subtraction
(we will switch to a different subtraction scheme momentarily).  Since the
divergences are linear the Feynman diagrams have poles at $D = 3$ but not at $D
= 4$.  In $MS$ (minimal subtraction) the coefficients of the operators
explicitly displayed in Eq.~(\ref{6}), are subtraction point independent and we
denote them by $\bar C_0^{(s)}$.  In this scheme the sum of bubble-type Feynman
diagrams gives
\begin{equation}\label{8}
{\cal A}^{(s)} = - \frac{ \bar C_0^{(s)}}{1 + iMp \bar C_0^{(s)}/4\pi}.
\end{equation}
Comparing Eq.~(\ref{8}) with eqs.~(\ref{1}) and~(\ref{2}) it is evident that
this corresponds to keeping only the scattering length term in the expansion of
$p\cot \delta^{(s)}$, (i.e., the first term of Eq.~(\ref{4})) and that
\begin{equation}\label{9}
\bar C_0^{(s)} = \frac{4\pi a^{(s)}}{M}.
\end{equation}
So in this subtraction scheme the coefficients $\bar C_0^{(s)}$ are very large
and also very different in the two channels.  However as $a^{(s)} \rightarrow
\infty, {\cal A}^{(s)} \rightarrow 4\pi i/Mp$ which is the same in both
channels.
This form for the scattering amplitudes is consistent with 
Wigner spin-isospin $SU(4)$ symmetry, and also with scale invariance.

In MS when $p> 1/a^{(s)}$ the terms in the perturbative series for the
scattering amplitude get larger and larger.  We would like to use a subtraction
scheme where the various Feynman diagrams in Fig. 1 are the same size as their
sum and where the symmetries that arise as $a^{(s)} \rightarrow \infty$ are
manifest at the level of the Lagrangian.  Examples of such subtraction schemes
are~\cite{kaplan1} PDS where poles at $D=3$ are also subtracted and the $OS$
momentum space
subtraction scheme.~\cite{weinberg1,mehen1}  In these schemes the
coefficients are subtraction
point dependent, $C_0^{(s)} = C_0^{(s)} (\mu)$, and the sum of bubble diagrams
gives
\begin{equation}\label{10}
{\cal A}^{(s)} = - \frac{C_0^{(s)} (\mu)}{1 + M (\mu + ip) C_0^{(s)}
(\mu)/4\pi}.
\end{equation}
This still corresponds to keeping just the scattering length, and is the
leading term in Eq.~(\ref{4}).  But now
\begin{equation}\label{11}
C_0^{(s)} (\mu) = - \frac{4\pi}{M} \frac{1}{\mu - 1/a^{(s)}},
\end{equation}
which as $a^{(s)} \rightarrow \infty$ becomes $C_0^{(s)} (\mu) = - 4\pi/M\mu$.
In this limit, the coefficients are the same in both channels and with
$\mu \sim p$ each term in sum of bubble type Feynman diagrams in Fig. 1 is the
same size as the sum itself.

The operators with coefficients $C_0^{(s)}$ are nonrenormalizable dimension six
operators.  Naively they are irrelevant operators and at low momentum can be
treated in perturbation theory.  However as $a^{(s)} \rightarrow \infty$ the
coefficients $C_0^{(s)} (\mu)$ flow to a nontrivial fixed
point~\cite{kaplan1,weinberg1} where $\mu d
[\mu C_0^{(s)} (\mu)]/d\mu = 0$.  For large $a^{(s)}$ the power counting is
controlled by this fixed point and the leading contribution to the $NN$
scattering amplitude comes from treating $C_0^{(s)}$ nonperturbatively.  It is
straightforward to show that in PDS or OS the coefficients of $S$-wave
operators with $2n$ spatial derivatives scale as,~\cite{kaplan1}
\begin{equation}\label{12}
C_{2n}^{(s)} (\mu) \sim \frac{4\pi}{M\Lambda^n\mu^{n+1}},
\end{equation}
for $\mu \gg |1/a^{(s)}|$.  With $\mu \sim p, C_{2n}^{(s)} (\mu) p^{2n} \sim
p^{n-1}$ and the two body operators with derivatives can be treated
perturbatively.  In a non-relativistic theory a loop integration $\int d^4 q =
\int dq^0 d^3 q \sim {\cal O} (p^5)$ (since the $dq^0$ integration is of
order $p^2$ and the $d^3q$ is order $p^3$) and the nucleon propagator $i/(p^0 -
p^2/2M + i\epsilon) \sim {\cal O} (p^{-2})$.  Consequently each loop gives a
factor $p$ plus whatever factors of $p$ are associated with the vertices.  The
power counting~\cite{kaplan1,kolck} is now evident.  The leading order $(LO)$
contribution
${\cal A}_{-1}^{(s)}$ comes from $C_0^{(s)}$ treated nonperturbatively, the
next to leading order $(NLO)$ contribution ${\cal A}_0^{(s)}$ comes from
$C_0^{(s)}$ treated nonperturbatively and $C_2^{(s)}$ inserted once, the
next-to-next to leading order $(N^2 LO)$ contribution comes from $C_0^{(s)}$
treated nonperturbatively, $C_2^{(s)}$ inserted twice or $C_4^{(s)}$ inserted
once, etc.

With the pions integrated out the effective field theory expansion applied to
$NN$ scattering reproduces Eq.~(\ref{4}) and has no more content than the
momentum expansion of $p\cot\delta^{(s)}$.  However, even with the pions
integrated out one can couple photons or $W$ and $Z$ gauge bosons to the
nucleons.  The relative importance of operators containing these fields depends
on their renormalization group scaling near the fixed point.

In the two nucleon sector predictions based on the effective field theory
without pions are similar to those made by effective range theory.~\cite{bethe}
 However the
effective field theory approach has a number of advantages.  Predictions based
on effective range theory are only valid to a given order in the $p/\Lambda$
expansion.  In the effective field theory new two-body operators containing the
gauge fields arise which spoil the predictions of effective range theory.  For
the thermal neutron capture cross section, $\sigma(n + p \rightarrow d +
\gamma)$, this occurs at $NLO$ while for the deuteron matter (charge) radius
$<r_m>$ this doesn't occur until $N^3LO$.  This explains why the effective
range theory prediction for $\sigma(n + p \rightarrow d + \gamma)$ is off by
10\% while the effective range theory prediction for $<r_m>$ is accurate to
better than a percent.\cite{chen}  For these static deuteron properties the
relevant
momentum in the $p/\Lambda$ expansion is set by the deuteron binding energy,
i.e., $p \sim 40$ MeV. Another useful aspect of the effective field theory
formalism is that it is straightforward to include relativistic corrections.

As $a^{(s)} \rightarrow \infty, {\cal L}_2 \rightarrow - (2\pi/M\mu)
(N^\dagger N)^2 + \ldots$, where the ellipses denote two body operators with
derivatives.  In this limit the leading one and two body terms are invariant
under the following symmetries:~\cite{mehen2}
\begin{description}
\item{(i.)} Wigner Symmetry~\cite{wigner}\\
Under infinitesimal Wigner symmetry $SU(4)$ transformations
\begin{equation}\label{13}
\delta N = i \alpha_{\mu\nu} \sigma^\mu \tau^\nu N,
\end{equation}
where $\sigma^\mu = (1,\mbox{\boldmath $\sigma$})$ and $\tau^\mu = (1,
\mbox{\boldmath $\tau$})$ with $\mu = 0,1,2,3$ and repeated indices summed.
The symmetry group corresponding to Eq.~(\ref{13}) is actually $SU(4)\times
U(1)$, with $\alpha_{00}$ the group parameter for the additional baryon-number
$U(1)$.  Associated with this symmetry are the conserved charges,
\begin{equation}\label{14}
Q^{\mu\nu} = \int d^3 x N^\dagger \sigma^\mu\tau^\nu N.
\end{equation}
The two body terms with derivatives are not invariant under Wigner symmetry
even if $a^{(s)} \rightarrow \infty$.  Hence in the two body sector the
violations of Wigner symmetry go as, $(1/[a^{(^1S_0)}p] - 1/[a^{(^3S_1)} p])$
and $p/\Lambda$.  Wigner symmetry will not be a good approximation if the
momentum $p$ is too low or if it is too large.

Wigner symmetry is relevant for nuclei with many
nucleons.~\cite{group}
It is not difficult to see that the higher body terms with no derivatives are
automatically invariant under Wigner symmetry.  Since these contact terms are
antisymmetric in the nucleon fields $N$ and in the hermitian conjugates
$N^\dagger$, contact terms without derivatives cannot occur for five body
operators and higher.  The nucleons $N$ are in the ${\bf 4}$ of $SU(4)$ and the
$N^\dagger$'s are in the $\bar{\bf 4}$.  Four nucleons combined anti-symmetrically
are an $SU(4)$ singlet and so the four-body terms are  invariant under $SU(4)$.
 The three body terms transform as $\bar{\bf 4} \otimes {\bf 4} = {\bf 1}
\otimes {\bf 15}$.  However the operators in the ${\bf 15}$ are not invariant
under the total spin or isospin $SU(2)$ subgroups of $SU(4)$.  Hence the
allowed three body terms are also invariant under $SU(4)$ Wigner symmetry.

A complete extension of the general fixed point power counting to the higher
body terms has not been made.  However there has been considerable recent
progress.~\cite{kolck}  This work indicates that the 3-body term with no
derivatives is
leading order (i.e., as important as effects coming from $C_0^{(s)}$).
\item{(ii.)}  Scale Invariance\\
The leading one and two body terms are invariant under the scale transformation
$N(t,{\bf x}) \rightarrow N' (t,{\bf x})$ and $\mu \rightarrow \mu'$ where
\begin{equation}\label{15}
N' (t,{\bf x}) = \lambda^{-3/2} N (t/\lambda^2, {\bf x}/\lambda),
\end{equation}
\begin{equation}\label{16}
\mu' = \mu/\lambda.
\end{equation}
Note that Eq.~(\ref{15}) corresponds to $N'(t', {\bf x}') = \lambda^{-3/2} N(t,
{\bf x})$ with ${\bf x}' = \lambda {\bf x}$ and $t' = \lambda^2 t$.  The
different scaling of space and time coordinates is dictated by invariance of
the leading one-body terms in the Lagrange density.
\end{description}

\section{Including Pions}

With pions included the power counting is taken to be in powers of
$Q/\Lambda_{NN}$ where $p\sim m_\pi \sim Q$.  A subscript $NN$ has been put on
$\Lambda$ as a reminder that the expansion should work better if the pions are
included as explicit fields, i.e., we expect that $\Lambda_{NN} > \Lambda$.
Potential pion exchange arises from the term
\begin{equation}\label{17}
{\cal L}_{{\rm int}} = - \frac{g_A}{\sqrt{2} f_\pi} \nabla^i \pi^j N^\dagger
\sigma^i \tau^j N,
\end{equation}
where $g_A \simeq 1.25$ is the axial coupling and $f_\pi \simeq 131$ MeV is the
pion decay constant.  Exchange of a potential pion between nucleons is order
$Q^0$ (the two factors of $Q$ from the vertices cancel the $1/Q^2$ from the
pion propagator).  This is the same size as the two body contact terms with two
derivatives and consequently pion exchange 
can be treated perturbatively.  Including pion
exchanges without the two derivative two body contact terms is not a systematic
improvement and is no better (from a power counting perspective) than just
including the effects of $C_0^{(s)}$.  Note that this power counting is very
different from the one originally proposed by
Weinberg~\cite{weinberg1,weinberg2} where the leading contribution came
from treating both potential pion
exchange and $C_0^{(s)}$ nonperturbatively. The effects of two
body terms with derivatives and insertions of the light quark mass
matrix were considered subdominant. 

Weinberg's power counting treats the nucleon mass $M$ as large and
$MQ \sim {\cal O}(1)$. It assumes that factors of $M$ only arise
from the loop integrations. In a toy model where perturbative matching
between the full relativistic theory and the nonrelativistic effective
theory can be explicitly performed Luke and Manohar~ {\cite{luke}} found
that the two body local operators in the effective nonrelativistic theory
have coefficients that contain factors of M. This is the origin of the problem
with Weinberg's power counting.

\begin{figure}[t!]
\centerline{\psfig{figure=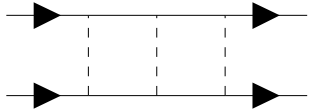,height=3.0cm,rwidth=16cm}}
\caption[1]{Contribution to $^3S_1$ scattering from
three potential pion exchange.  The dashed lines denote potential pion
exchange.}
\label{fig_ct0pi}
\end{figure}
In the $^3S_1$
channel  the Feynman diagram shown in Fig. 2 with three potential pion
exchanges is logarithmically divergent.  Neglecting the pion mass it gives a
contribution to ${\cal A}^{(^3S_1)}$ of order
\begin{equation}\label{18}
\left(\frac{4\pi}{M}\right) \left(\frac{Mg_A^2}{8\pi f_\pi^2}\right)^3 p^2 [\ln
\mu^2 + K],
\end{equation}
where $K$ is a constant.  The $\mu$-dependence above is cancelled by the
$\mu$-dependence of $C_2^{(^3S_1)}$.  There is no point to including this
Feynman diagram without including the effects of the two body $^3S_1$ operator
with 2-derivatives.  Eq.~(\ref{18}) is an $N^3LO$ contribution.  With the pions
included a single insertion of $C_2^{(s)}$ is not just $NLO$ it contributes at
higher levels at the $Q$ expansion as well.  For that reason $C_2^{(s)}$ and
the other contact term coefficients are sometimes written as a sum $C_2^{(s)} =
\sum_{a = 1}^\infty C_{2,a}^{(s)}$ where $C_{2,1}^{(s)}$ gives the $NLO$
contribution, etc.  When this is done predictions for physical quantities are
exactly $\mu$ independent, at each order in the $Q$ expansion.  If $C_2^{(s)}$
is not expanded in this way then predictions at a given order on the $Q$
expansion have some subtraction point dependence, which is higher order in the
$Q$ expansion.

\begin{figure}[t!]
\centerline{\psfig{figure=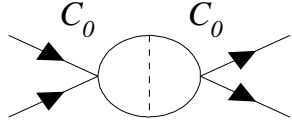,height=3.0cm,rwidth=16cm}}
\caption[1]{Contribution to $NN$ scattering that renormalizes $D_2^{(s)}$.}
\label{fig_ct1pi}
\end{figure}
There are two body $S$-wave contact terms with no derivatives but with an
insertion of the light quark mass matrix,
\begin{equation}\label{19}
m_q = \left(\begin{array}{cc}
m_u & 0\\
0 & m_d \end{array} \right).
\end{equation}
Since $m_\pi^2 \propto (m_u + m_d)$ an insertion of $m_q$ counts as two powers
of $Q$ and the coefficients of these operators $D_2^{(s)}$ scale with $\mu$ in
the same way as the coefficients $C_2^{(s)}$.  At $NLO$ they must also be
included.  The Feynman diagram in Fig. 3 is logaritmically divergent and it
gives a contribution to the $^1S_0$ and $^3S_1$ scattering amplitudes of order
\begin{equation}\label{20}
\left(\frac{4\pi}{M}\right) \left(\frac{g_A^2 M}{8\pi f_\pi^2}\right)
\left(\frac{C_0^{(s)} M}{4\pi}\right)^2 m_\pi^2 [\ln \mu^2 + K],
\end{equation}
where $K$ is a constant.  The $\mu$ dependence here is cancelled by that of the
coefficients $D_2^{(s)}$.  Including one pion exchange without the
effects of the two body terms with one insertion of the quark mass matrix does
not systematically improve the theoretical prediction for the $NN$ scattering
amplitude.

If a momentum cutoff regulator is used instead of dimensional regularization
then including pion exchange without the two body contact operators that have
an insertion of the quark mass matrix results in a cutoff dependent amplitude
${\cal A}^{(s)}$. It is possible in the $^1S_0$ channel to sum to all orders
potential pion exchange and when this is done the cutoff
dependence does not become subdominant~\cite{kaplan2} 
(compared with the finite cut off
independent parts of pion exchange).  The effects of local four
nucleon (i.e., two body) operators with an insertion of the quark mass matrix
cannot be viewed as less important than the effects of pion exchange.

The conventional explanation for the discrepancy between the
prediction of effective range theory for the thermal neutron capture cross
section $\sigma (n + p \rightarrow d + \gamma$) and its experimental value is
meson exchange currents,~\cite{riska} which roughly speaking are the
contribution of Feynman diagrams where the photon couples to a potential pion.
In the effective field theory approach with the pions included this discrepancy
is made up~\cite{savage} (at least partly) by the $NLO$ contribution which
involves both meson exchange current Feynman diagrams and the contribution of a
local two body operator involving the magnetic field.

\section{An Application of Wigner Symmetry}

Potential pions have $k^0 \sim {\bf k}^2/M$ while radiation pions have $k^0 \sim
\sqrt{{\bf k}^2 + m_\pi^2}$.  The coupling of the radiation pions to the
nucleons is done by performing a multipole expansion on Eq.~(\ref{17}).  At
leading order this
amounts to evaluating the pion field in Eq.~(\ref{17}) at the space time point
$(t,{\bf x}) = (t, {\bf 0})$.  Hence, for radiation pions the term in the
action
corresponding to Eq.~(\ref{17}) is
\begin{equation}\label{21}
S_{int} = - \frac{g_A}{\sqrt{2} f} \int dt (\nabla^i \pi^j)|_{{\bf x} = 0}
Q^{ij},
\end{equation}
where $Q^{ij}$ are the charges of Wigner symmetry in Eq.~(\ref{14}).  In the
limit $a^{(s)} \rightarrow  \infty$ these charges are conserved and the
$Q^{ij}$'s are time independent.  Hence, as $a^{(s)} \rightarrow \infty$ only
the $k^0 = 0$ mode of the pion couples in Eq.~(\ref{21}).  This is incompatible
with the radiation pion condition, $k^0 \sim \sqrt{{\bf k}^2 + m_\pi^2}$.
Hence the leading contribution from radiation pions is suppressed by
$1/a^{(^1S_0)} - 1/a^{(^3S_1)}$.  In a recent paper~\cite{mehen3} Mehen and
Stewart found
this suppression by an explicit calculation of the leading radiation pion
contribution to the $NN$ scattering amplitudes, ${\cal A}^{(s)}$.  It
involved a cancellation between different Feynman diagrams.

\section{Outlook}

Effective field theory methods are a viable model independent approach to
the physics of the two nucleon sector. The power counting is slightly
unusual due to the large $S$-wave $NN$ scattering lengths. 
This approach is useful up to a center of mass momentum around
$200~{\rm MeV}$, however, the expansion parameter at
such a momentum is probably not much
smaller than ${1 \over 2}$. It seems likely that for many quantities
calculations at $N^2LO$ will reach the same precision as conventional
potential model approaches, however, with such a large expansion parameter
there are likely to be some failures.

Extension of the effective field theory approach to the three nucleon sector
is underway. Several theoretical issues remain to be resolved before
there is a complete power counting, but recent progress in this area is
very encouraging. 

The holy grail of this field is the application of these effective field theory
methods to nuclear matter. We are still a long way from having the theoretical
tools to tackle this problem and even with these tools the Fermi momentum
associated with nuclear density may be too large for a $Q$ expansion
to be useful. However, given the importance of understanding the
properties of nuclear matter continuing to develop 
the effective field theory approach is very worthwhile.

This work was supported in part by the Department of Energy under grant
number DE-FG03-92-ER 40701.


\section*{References}
\begin{thebibliography}{99}

\bibitem{burcham}W.E. Burcham, {\em Elements of Nuclear Physics}, John Wiley
and Sons Inc., (1979).

\bibitem{kaplan1}D.B. Kaplan, M.J. Savage and M.B. Wise, {\em Phys. Lett.} {\bf
B424}, 390 (1998); {\em Nucl. Phys.} {\bf B534}, 329 (1998).

\bibitem{weinberg1}S. Weinberg, {\em Nucl. Phys.} {\bf B363}, 3 (1991).

\bibitem{mehen1}T. Mehen and I.W. Stewart, {\em Phys. Lett.} {\bf B445}, 378
(1999); T. Mehen and I.W. Stewart, nucl-th/9809095; J. Gegelia,
nucl-th/9802038.

\bibitem{kolck}U. van Kolck, hep-ph/971222 and {\em Nucl. Phys.} {\bf A645},
273 (1999).

\bibitem{bethe}H.A. Bethe, {\em Phys. Rev.} {\bf 76}, 38 (1949); H.A. Bethe and
C. Longmire, {\em Phys. Rev.} {\bf 77}, 647 (1950).

\bibitem{chen}J.-W. Chen, G. Rupak and M.J. Savage, nucl-th/9962056.

\bibitem{mehen2}T. Mehen, I.W. Stewart and M.B. Wise, hep-ph/9902370.

\bibitem{wigner}E. Wigner, {\em Phys. Rev.} {\bf 51}, 106 (1937); ibid {\bf
56}, 519 (1939).

\bibitem{group}{\em Group Symmetries in Nuclear Structure}, J.C. Parikh, Plenum
press, (1978); T.W. Donnelly and G.E. Walker, {\em Ann. of Phys.} {\bf 60}, 209
(1970); J.D. Walecka, {\em Theoretical Nuclear Physics and Subnuclear Physics},
Oxford University Press (1995);  J.P. Elliott, {\em Isospin in Nuclear
Physics}, ed. D.H. Wilkinson, p73, North Holland Publishing (1969);  P. Vogel
and M.R. Zirnbauer, {\em Phys. Rev. Lett.} {\bf 57}, 3148 (1986); 
 P. Vogel and W.E.
Ormand, {\em Phys. Rev.} {\bf C47}, 623 (1993);  Y.V. Gaponov, N.B. Shulgina,
and D.M. Vladimirov, {\em Nucl. Phys.} {\bf A391}, 93 (1982).

\bibitem{bedaque}P.F. Bedaque and U. van Kolck, {\em Phys. Lett.} {\bf B428},
221 (1998);  P.F. Bedaque, H.W. Hammer and U. van Kolck, {\em Phys. Rev.} {\bf
C58}, R641 (1998); {\em Phys. Lett.} {\bf 82}, 463 (1999); {\em Nucl. Phys.}
{\bf A646}, 444 (1999).

\bibitem{weinberg2}S. Weinberg, {\em Phys. Lett.} {\bf B25}, 288 (1990).

\bibitem{luke}M. Luke and A. V. Manohar, {\em Phys. Rev.} {\bf D55}, 4129
(1997).

\bibitem{kaplan2}D.B. Kaplan, M.J. Savage and M.B. Wise, {\em Nucl. Phys.} {\bf
B478}, 629 (1996).

\bibitem{riska}D.O. Riska and G.E. Brown, {\em Phys. Lett.} {\bf B38}, 193
(1972).

\bibitem{savage}M.J. Savage, K.A. Scaldeferri and M.B. Wise, nucl-th/9811029.

\bibitem{mehen3}T. Mehen and I. Stewart, nucl-th/9901064.




\end{thebibliography}
\end{document}